\documentclass[
 reprint,
 amsmath,
 amssymb,
 aps, onecolumn
]{revtex4-2}

\usepackage{graphicx}
\usepackage{dcolumn}
\usepackage{bm}
\usepackage{color, float}

\DeclareMathAlphabet{\pazocal}{OMS}{zplm}{m}{n}

\begin{document}

\title{Stability of asymptotically flat (2+1)-dimensional black holes with Gauss-Bonnet corrections \\
{\small \rm Prepared for the 25th RAGtime workshop, 27 November -- 1 December, Opava}
}

\author{Milena Skvortsova}
\email{milenas577@mail.ru}
\affiliation{Peoples' Friendship University of Russia (RUDN University), 6 Miklukho-Maklaya Street, Moscow, 117198, Russia}


\begin{abstract}
Using the integration of wave equation in time-domain we show that scalar field perturbations around the $(2+1)$-dimensional asymptotically flat black hole with Gauss-Bonnet corrections is dynamically stable even for the near extreme values of the coupling constant.
\end{abstract}

\maketitle

\section{Introduction}\label{intro}

Black holes have been observed in the gravitational  \cite{LIGOScientific:2016aoc} and electromagnetic \cite{EventHorizonTelescope:2019dse,Goddi:2016qax} spectra, while ongoing projects promise a great extension of the range of observations \cite{LISACosmologyWorkingGroup:2022jok}.
Quasinormal modes \cite{Berti:2009kk,Nollert:1999ji,Kokkotas:1999bd,Konoplya:2011qq} represent the characteristic oscillations that black holes undergo when perturbed, and they serve as a valuable tool for probing the black holes' fundamental properties.  Our study explores the intricate interplay between black hole's quasinormal modes and their (in)stability. Usually, the strict mathematical proof of stability is a very difficult problem (see, for instance, \cite{Beyer:2011py}). A clear evidence of the linear stability or instability is provided by the analysis of the quasinormal spectrum. If all the quasinormal modes are damped, the black hole is stable, while if there exists at least one unboundedly growing mode, it signifies the onset of instability. Therefore, the instability was extensively studied with the help of quasinormal modes spectra \cite{Takahashi:2010gz,Ishihara:2008re,Kodama:2009bf,Dyatlov:2010hq}. 

The particular black hole metric we will study here is the $(2+1)$-dimensional black hole obtained as a regularization \cite{Glavan:2019inb} of the Einstein-Gauss-Bonnet equations of motion in \cite{Konoplya:2020ibi} and further studied and generalized in \cite{Hennigar:2020fkv,Hennigar:2020drx}. Although the straightforward  regularization in \cite{Glavan:2019inb} did not form a consistent theory in four dimensions, the black hole solution obtained within  such a naive approach were also solutions of the well-defined theory formulated in \cite{Aoki:2020lig}.

Quasinormal modes of three-dimensional asymptotically AdS (BTZ) spacetimes were extensively studied \cite{Cardoso:2001hn,Konoplya:2004ik,Fontana:2023dix}, because of their importance in the AdS/CFT correspondence \cite{Birmingham:2001pj}. However, no such analysis has been done so far for the $(2+1)$-dimensional asymptotically flat spacetimes, to the best of our knowledge. 
Here we will show that the scalar field perturbations around the $(2+1)$-dimensional asymptotically flat black holes  decay with time even at the near extreme values of the Gauss-Bonnet coupling.

\section{Black hole metrics and wavelike equations}\label{sec.Bardeen spacetime and the wavelike equations}

The metric of the (2+1)-dimensional black hole has the following form
\begin{equation}\label{spherical}
\mathrm{d}s^2 = -f(r)\mathrm{d} t^2 + f(r)^{-1} \mathrm{d} r^2 + r^2  \mathrm{d} x ^2,
\end{equation}
where the metric function is 
\begin{equation}
f(r) =1-\frac{r^2}{2 \alpha} \left(-1 + \sqrt{\frac{4 \alpha \left(\Lambda 
   \left(r^2-r_H^2\right)+\frac{\alpha}{r_H^2}+1\right)}{r^2}+1}\right).
\end{equation}
Here, $r_H$ is the radius of the event horizon, $\Lambda$ is the cosmological constant, $\alpha$ is the Gauss-Bonnet coupling constant.   
When $1+ 2 \alpha/r_{H}^2 >0$ and the coupling constant $\alpha <0$ at $\Lambda =0$
the metric is asymptotically flat and perturbative in $\alpha$. When the cosmological constant is negative, the metric is reduced to the BTZ one \cite{Banados:1992wn} up to the redefinition of constants. 
The general-covariant Klein-Gordon equation
\begin{equation}
\frac{1}{\sqrt{-g}}\partial_\mu \left(\sqrt{-g}g^{\mu \nu}\partial_\nu\Phi\right) =0
\end{equation}
can be reduced to the wave-like form 
\begin{equation}\label{wave-equation}
\dfrac{d^2 \Psi}{dr_*^2}+(\omega^2-V(r))\Psi=0,
\end{equation}
where the ``tortoise coordinate'' $r_*$ has the form:
\begin{equation}
dr_*\equiv\frac{dr}{f(r)},
\end{equation}
and the effective potential is
\begin{equation}\label{potentialScalar}
V(r)=f(r) \left(\frac{k^2}{r^2}+\frac{f'(r)}{2 r} - \frac{f(r)}{4 r^2} \right)
\end{equation}
Here  $k$ is the multipole number.

\section{Evolution of perturbations}\label{geneqcond}

Quasinormal modes of asymptotically flat black holes are eigenvalues of the above wave-like equation satisfying particular boundary conditions: the waves must be purely outgoing at infinity and purely incoming at the event horizon. 
For analysis of the evolution of perturbations in time and consideration of contribution of all the overtones, so that the instability, if any, could be detected, we use the time-domain integration method \cite{Gundlach:1993tp}, which was used in numerous works (for instance, \cite{Konoplya:2007yy,Churilova:2019cyt,Bronnikov:2019sbx}). It shows a good concordance with more accurate methods.  The essence of this method is integration of the wave equation in the framework of the null-cone coordinates $u=t-r_*$, $v=t+r_*$, using the following discretization \cite{Gundlach:1993tp}, 
\begin{equation}\label{Discretization}
\Psi\left(N\right)=\Psi\left(W\right)+\Psi\left(E\right)-\Psi\left(S\right) - \Delta^2V\left(S\right)\frac{\Psi\left(W\right)+\Psi\left(E\right)}{4}+{\cal O}\left(\Delta^4\right).
\end{equation}
Here, the points are: $N\equiv\left(u+\Delta,v+\Delta\right)$, $W\equiv\left(u+\Delta,v\right)$, $E\equiv\left(u,v+\Delta\right)$, $S\equiv\left(u,v\right)$, and the Gaussian impinging wave package are given on the null surfaces $u=u_0$ and $v=v_0$. 

For checking the results obtained by the time-domain integration at $k>0$ we used the 6th order WKB method with Pad\'e approximants \cite{Konoplya:2019hlu,Matyjasek:2017psv,Konoplya:2003ii}.

\begin{figure}[H]
	\begin{center}
		\begin{tabular}{cc}
			\includegraphics[width=3in]{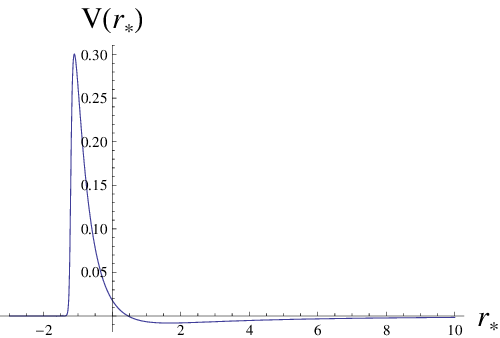} 
    \includegraphics[width=3in]{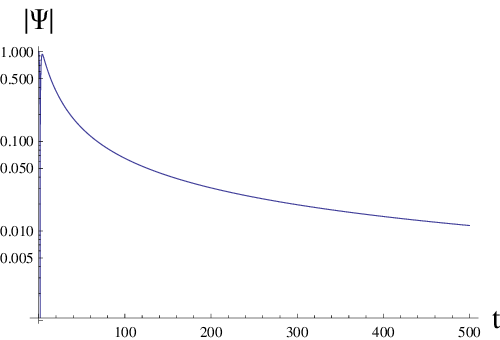}\\
		\end{tabular}
		\caption{Left: Effective potential as a function of $r_{*}$. Right: Time-domain profile of perturbations. Here we have $k=0$, $\Lambda=0$, $\alpha=-0.49$. The Prony method suggests that the dominant modes are non-oscillatory $\omega = - 0.001899 i$}
		\label{fig:Potential}
	\end{center}
\end{figure}

\begin{figure}[H]
	\begin{center}
		\begin{tabular}{cc}
			\includegraphics[width=3in]{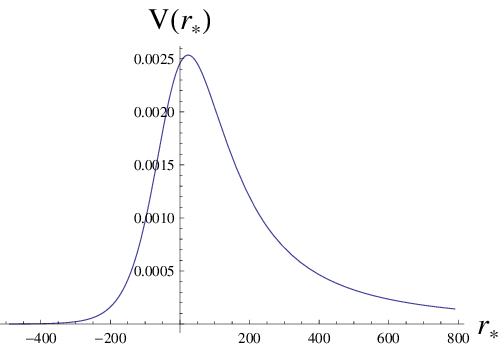} 
    \includegraphics[width=3in]{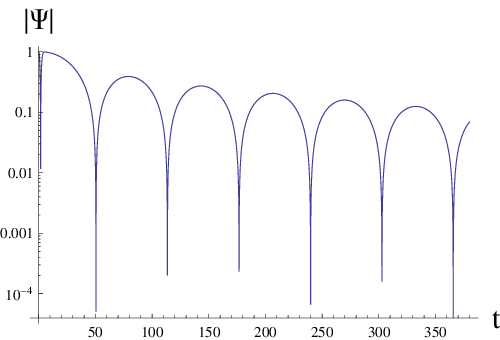}\\
		\end{tabular}
		\caption{Left: Effective potential as a function of $r_{*}$. Right: Time-domain profile of perturbations. Here we have  $k=1$, $\Lambda=0$, $\alpha=-0.01$. The Prony method gives $\omega = 0.050145 - 0.003598 i$, while the 6th order WKB method with Pade approximants  gives $\omega =0.050155-0.003579 i$}
		\label{fig:Potential2}
	\end{center}
\end{figure}

Time-domain profiles show decaying profiles not only for small  $\alpha$ (see Fig. \ref{fig:Potential2}), but also for the extreme $\alpha$, as in Fig. \ref{fig:Potential}. If the instability appeared, it would be governed by the non-oscillatory, i.e. purely imaginary, growing mode, as was proved in \cite{Konoplya:2008yy} for generic static backgrounds. 

\section{Conclusions}\label{conclus}

Here we have shown that asymptotically flat $(2+1)$-dimensional black holes with Gauss-Bonnet corrections are stable even at the near extreme values of the Gauss-Bonnet coupling. The full parametric range of stability and the other types of asymptotics could be studied in further research. Thus, allowing for a non-zero cosmological constant may bring instabilities to the system \cite{Zhu:2014sya,Konoplya:2014lha}.

\acknowledgments{
The author acknowledges Dr. R. A. Konoplya for fruitful discussions, and for careful reading of the manuscript. This work was supported by RUDN University research project FSSF-2023-0003.}

\bibliographystyle{apsrev4-2}
\bibliography{ragsamp}

\begin{thebibliography}{33}%
\makeatletter
\providecommand \@ifxundefined [1]{%
 \@ifx{#1\undefined}
}%
\providecommand \@ifnum [1]{%
 \ifnum #1\expandafter \@firstoftwo
 \else \expandafter \@secondoftwo
 \fi
}%
\providecommand \@ifx [1]{%
 \ifx #1\expandafter \@firstoftwo
 \else \expandafter \@secondoftwo
 \fi
}%
\providecommand \natexlab [1]{#1}%
\providecommand \enquote  [1]{``#1''}%
\providecommand \bibnamefont  [1]{#1}%
\providecommand \bibfnamefont [1]{#1}%
\providecommand \citenamefont [1]{#1}%
\providecommand \href@noop [0]{\@secondoftwo}%
\providecommand \href [0]{\begingroup \@sanitize@url \@href}%
\providecommand \@href[1]{\@@startlink{#1}\@@href}%
\providecommand \@@href[1]{\endgroup#1\@@endlink}%
\providecommand \@sanitize@url [0]{\catcode `\\12\catcode `\$12\catcode
  `\&12\catcode `\#12\catcode `\^12\catcode `\_12\catcode `\%12\relax}%
\providecommand \@@startlink[1]{}%
\providecommand \@@endlink[0]{}%
\providecommand \url  [0]{\begingroup\@sanitize@url \@url }%
\providecommand \@url [1]{\endgroup\@href {#1}{\urlprefix }}%
\providecommand \urlprefix  [0]{URL }%
\providecommand \Eprint [0]{\href }%
\providecommand \doibase [0]{https://doi.org/}%
\providecommand \selectlanguage [0]{\@gobble}%
\providecommand \bibinfo  [0]{\@secondoftwo}%
\providecommand \bibfield  [0]{\@secondoftwo}%
\providecommand \translation [1]{[#1]}%
\providecommand \BibitemOpen [0]{}%
\providecommand \bibitemStop [0]{}%
\providecommand \bibitemNoStop [0]{.\EOS\space}%
\providecommand \EOS [0]{\spacefactor3000\relax}%
\providecommand \BibitemShut  [1]{\csname bibitem#1\endcsname}%
\let\auto@bib@innerbib\@empty
\bibitem [{\citenamefont {Abbott}\ \emph {et~al.}(2016)\citenamefont {Abbott}
  \emph {et~al.}}]{LIGOScientific:2016aoc}%
  \BibitemOpen
  \bibfield  {author} {\bibinfo {author} {\bibfnamefont {B.~P.}\ \bibnamefont
  {Abbott}} \emph {et~al.} (\bibinfo {collaboration} {LIGO Scientific,
  Virgo}),\ }\href {https://doi.org/10.1103/PhysRevLett.116.061102} {\bibfield
  {journal} {\bibinfo  {journal} {Phys. Rev. Lett.}\ }\textbf {\bibinfo
  {volume} {116}},\ \bibinfo {pages} {061102} (\bibinfo {year} {2016})},\
  \Eprint {https://arxiv.org/abs/1602.03837} {arXiv:1602.03837 [gr-qc]}
  \BibitemShut {NoStop}%
\bibitem [{\citenamefont {Akiyama}\ \emph {et~al.}(2019)\citenamefont {Akiyama}
  \emph {et~al.}}]{EventHorizonTelescope:2019dse}%
  \BibitemOpen
  \bibfield  {author} {\bibinfo {author} {\bibfnamefont {K.}~\bibnamefont
  {Akiyama}} \emph {et~al.} (\bibinfo {collaboration} {Event Horizon
  Telescope}),\ }\href {https://doi.org/10.3847/2041-8213/ab0ec7} {\bibfield
  {journal} {\bibinfo  {journal} {Astrophys. J. Lett.}\ }\textbf {\bibinfo
  {volume} {875}},\ \bibinfo {pages} {L1} (\bibinfo {year} {2019})},\ \Eprint
  {https://arxiv.org/abs/1906.11238} {arXiv:1906.11238 [astro-ph.GA]}
  \BibitemShut {NoStop}%
\bibitem [{\citenamefont {Goddi}\ \emph {et~al.}(2016)\citenamefont {Goddi}
  \emph {et~al.}}]{Goddi:2016qax}%
  \BibitemOpen
  \bibfield  {author} {\bibinfo {author} {\bibfnamefont {C.}~\bibnamefont
  {Goddi}} \emph {et~al.},\ }\href {https://doi.org/10.1142/9789813226609_0046}
  {\bibfield  {journal} {\bibinfo  {journal} {Int. J. Mod. Phys. D}\ }\textbf
  {\bibinfo {volume} {26}},\ \bibinfo {pages} {1730001} (\bibinfo {year}
  {2016})},\ \Eprint {https://arxiv.org/abs/1606.08879} {arXiv:1606.08879
  [astro-ph.HE]} \BibitemShut {NoStop}%
\bibitem [{\citenamefont {Auclair}\ \emph {et~al.}(2023)\citenamefont {Auclair}
  \emph {et~al.}}]{LISACosmologyWorkingGroup:2022jok}%
  \BibitemOpen
  \bibfield  {author} {\bibinfo {author} {\bibfnamefont {P.}~\bibnamefont
  {Auclair}} \emph {et~al.} (\bibinfo {collaboration} {LISA Cosmology Working
  Group}),\ }\href {https://doi.org/10.1007/s41114-023-00045-2} {\bibfield
  {journal} {\bibinfo  {journal} {Living Rev. Rel.}\ }\textbf {\bibinfo
  {volume} {26}},\ \bibinfo {pages} {5} (\bibinfo {year} {2023})},\ \Eprint
  {https://arxiv.org/abs/2204.05434} {arXiv:2204.05434 [astro-ph.CO]}
  \BibitemShut {NoStop}%
\bibitem [{\citenamefont {Berti}\ \emph {et~al.}(2009)\citenamefont {Berti},
  \citenamefont {Cardoso},\ and\ \citenamefont {Starinets}}]{Berti:2009kk}%
  \BibitemOpen
  \bibfield  {author} {\bibinfo {author} {\bibfnamefont {E.}~\bibnamefont
  {Berti}}, \bibinfo {author} {\bibfnamefont {V.}~\bibnamefont {Cardoso}},\
  and\ \bibinfo {author} {\bibfnamefont {A.~O.}\ \bibnamefont {Starinets}},\
  }\href {https://doi.org/10.1088/0264-9381/26/16/163001} {\bibfield  {journal}
  {\bibinfo  {journal} {Class. Quant. Grav.}\ }\textbf {\bibinfo {volume}
  {26}},\ \bibinfo {pages} {163001} (\bibinfo {year} {2009})},\ \Eprint
  {https://arxiv.org/abs/0905.2975} {arXiv:0905.2975 [gr-qc]} \BibitemShut
  {NoStop}%
\bibitem [{\citenamefont {Nollert}(1999)}]{Nollert:1999ji}%
  \BibitemOpen
  \bibfield  {author} {\bibinfo {author} {\bibfnamefont {H.-P.}\ \bibnamefont
  {Nollert}},\ }\href {https://doi.org/10.1088/0264-9381/16/12/201} {\bibfield
  {journal} {\bibinfo  {journal} {Class. Quant. Grav.}\ }\textbf {\bibinfo
  {volume} {16}},\ \bibinfo {pages} {R159} (\bibinfo {year}
  {1999})}\BibitemShut {NoStop}%
\bibitem [{\citenamefont {Kokkotas}\ and\ \citenamefont
  {Schmidt}(1999)}]{Kokkotas:1999bd}%
  \BibitemOpen
  \bibfield  {author} {\bibinfo {author} {\bibfnamefont {K.~D.}\ \bibnamefont
  {Kokkotas}}\ and\ \bibinfo {author} {\bibfnamefont {B.~G.}\ \bibnamefont
  {Schmidt}},\ }\href {https://doi.org/10.12942/lrr-1999-2} {\bibfield
  {journal} {\bibinfo  {journal} {Living Rev. Rel.}\ }\textbf {\bibinfo
  {volume} {2}},\ \bibinfo {pages} {2} (\bibinfo {year} {1999})},\ \Eprint
  {https://arxiv.org/abs/gr-qc/9909058} {arXiv:gr-qc/9909058} \BibitemShut
  {NoStop}%
\bibitem [{\citenamefont {Konoplya}\ and\ \citenamefont
  {Zhidenko}(2011)}]{Konoplya:2011qq}%
  \BibitemOpen
  \bibfield  {author} {\bibinfo {author} {\bibfnamefont {R.~A.}\ \bibnamefont
  {Konoplya}}\ and\ \bibinfo {author} {\bibfnamefont {A.}~\bibnamefont
  {Zhidenko}},\ }\href {https://doi.org/10.1103/RevModPhys.83.793} {\bibfield
  {journal} {\bibinfo  {journal} {Rev. Mod. Phys.}\ }\textbf {\bibinfo {volume}
  {83}},\ \bibinfo {pages} {793} (\bibinfo {year} {2011})},\ \Eprint
  {https://arxiv.org/abs/1102.4014} {arXiv:1102.4014 [gr-qc]} \BibitemShut
  {NoStop}%
\bibitem [{\citenamefont {Beyer}(2011)}]{Beyer:2011py}%
  \BibitemOpen
  \bibfield  {author} {\bibinfo {author} {\bibfnamefont {H.~R.}\ \bibnamefont
  {Beyer}},\ }\href {https://doi.org/10.1063/1.3653840} {\bibfield  {journal}
  {\bibinfo  {journal} {J. Math. Phys.}\ }\textbf {\bibinfo {volume} {52}},\
  \bibinfo {pages} {102502} (\bibinfo {year} {2011})},\ \Eprint
  {https://arxiv.org/abs/1105.4956} {arXiv:1105.4956 [math-ph]} \BibitemShut
  {NoStop}%
\bibitem [{\citenamefont {Takahashi}\ and\ \citenamefont
  {Soda}(2010)}]{Takahashi:2010gz}%
  \BibitemOpen
  \bibfield  {author} {\bibinfo {author} {\bibfnamefont {T.}~\bibnamefont
  {Takahashi}}\ and\ \bibinfo {author} {\bibfnamefont {J.}~\bibnamefont
  {Soda}},\ }\href {https://doi.org/10.1143/PTP.124.711} {\bibfield  {journal}
  {\bibinfo  {journal} {Prog. Theor. Phys.}\ }\textbf {\bibinfo {volume}
  {124}},\ \bibinfo {pages} {711} (\bibinfo {year} {2010})},\ \Eprint
  {https://arxiv.org/abs/1008.1618} {arXiv:1008.1618 [gr-qc]} \BibitemShut
  {NoStop}%
\bibitem [{\citenamefont {Ishihara}\ \emph {et~al.}(2008)\citenamefont
  {Ishihara}, \citenamefont {Kimura}, \citenamefont {Konoplya}, \citenamefont
  {Murata}, \citenamefont {Soda},\ and\ \citenamefont
  {Zhidenko}}]{Ishihara:2008re}%
  \BibitemOpen
  \bibfield  {author} {\bibinfo {author} {\bibfnamefont {H.}~\bibnamefont
  {Ishihara}}, \bibinfo {author} {\bibfnamefont {M.}~\bibnamefont {Kimura}},
  \bibinfo {author} {\bibfnamefont {R.~A.}\ \bibnamefont {Konoplya}}, \bibinfo
  {author} {\bibfnamefont {K.}~\bibnamefont {Murata}}, \bibinfo {author}
  {\bibfnamefont {J.}~\bibnamefont {Soda}},\ and\ \bibinfo {author}
  {\bibfnamefont {A.}~\bibnamefont {Zhidenko}},\ }\href
  {https://doi.org/10.1103/PhysRevD.77.084019} {\bibfield  {journal} {\bibinfo
  {journal} {Phys. Rev. D}\ }\textbf {\bibinfo {volume} {77}},\ \bibinfo
  {pages} {084019} (\bibinfo {year} {2008})},\ \Eprint
  {https://arxiv.org/abs/0802.0655} {arXiv:0802.0655 [hep-th]} \BibitemShut
  {NoStop}%
\bibitem [{\citenamefont {Kodama}\ \emph {et~al.}(2010)\citenamefont {Kodama},
  \citenamefont {Konoplya},\ and\ \citenamefont {Zhidenko}}]{Kodama:2009bf}%
  \BibitemOpen
  \bibfield  {author} {\bibinfo {author} {\bibfnamefont {H.}~\bibnamefont
  {Kodama}}, \bibinfo {author} {\bibfnamefont {R.~A.}\ \bibnamefont
  {Konoplya}},\ and\ \bibinfo {author} {\bibfnamefont {A.}~\bibnamefont
  {Zhidenko}},\ }\href {https://doi.org/10.1103/PhysRevD.81.044007} {\bibfield
  {journal} {\bibinfo  {journal} {Phys. Rev. D}\ }\textbf {\bibinfo {volume}
  {81}},\ \bibinfo {pages} {044007} (\bibinfo {year} {2010})},\ \Eprint
  {https://arxiv.org/abs/0904.2154} {arXiv:0904.2154 [gr-qc]} \BibitemShut
  {NoStop}%
\bibitem [{\citenamefont {Dyatlov}(2011)}]{Dyatlov:2010hq}%
  \BibitemOpen
  \bibfield  {author} {\bibinfo {author} {\bibfnamefont {S.}~\bibnamefont
  {Dyatlov}},\ }\href {https://doi.org/10.1007/s00220-011-1286-x} {\bibfield
  {journal} {\bibinfo  {journal} {Commun. Math. Phys.}\ }\textbf {\bibinfo
  {volume} {306}},\ \bibinfo {pages} {119} (\bibinfo {year} {2011})},\ \Eprint
  {https://arxiv.org/abs/1003.6128} {arXiv:1003.6128 [math.AP]} \BibitemShut
  {NoStop}%
\bibitem [{\citenamefont {Glavan}\ and\ \citenamefont
  {Lin}(2020)}]{Glavan:2019inb}%
  \BibitemOpen
  \bibfield  {author} {\bibinfo {author} {\bibfnamefont {D.}~\bibnamefont
  {Glavan}}\ and\ \bibinfo {author} {\bibfnamefont {C.}~\bibnamefont {Lin}},\
  }\href {https://doi.org/10.1103/PhysRevLett.124.081301} {\bibfield  {journal}
  {\bibinfo  {journal} {Phys. Rev. Lett.}\ }\textbf {\bibinfo {volume} {124}},\
  \bibinfo {pages} {081301} (\bibinfo {year} {2020})},\ \Eprint
  {https://arxiv.org/abs/1905.03601} {arXiv:1905.03601 [gr-qc]} \BibitemShut
  {NoStop}%
\bibitem [{\citenamefont {Konoplya}\ and\ \citenamefont
  {Zhidenko}(2020)}]{Konoplya:2020ibi}%
  \BibitemOpen
  \bibfield  {author} {\bibinfo {author} {\bibfnamefont {R.~A.}\ \bibnamefont
  {Konoplya}}\ and\ \bibinfo {author} {\bibfnamefont {A.}~\bibnamefont
  {Zhidenko}},\ }\href {https://doi.org/10.1103/PhysRevD.102.064004} {\bibfield
   {journal} {\bibinfo  {journal} {Phys. Rev. D}\ }\textbf {\bibinfo {volume}
  {102}},\ \bibinfo {pages} {064004} (\bibinfo {year} {2020})},\ \Eprint
  {https://arxiv.org/abs/2003.12171} {arXiv:2003.12171 [gr-qc]} \BibitemShut
  {NoStop}%
\bibitem [{\citenamefont {Hennigar}\ \emph {et~al.}(2020)\citenamefont
  {Hennigar}, \citenamefont {Kubiznak}, \citenamefont {Mann},\ and\
  \citenamefont {Pollack}}]{Hennigar:2020fkv}%
  \BibitemOpen
  \bibfield  {author} {\bibinfo {author} {\bibfnamefont {R.~A.}\ \bibnamefont
  {Hennigar}}, \bibinfo {author} {\bibfnamefont {D.}~\bibnamefont {Kubiznak}},
  \bibinfo {author} {\bibfnamefont {R.~B.}\ \bibnamefont {Mann}},\ and\
  \bibinfo {author} {\bibfnamefont {C.}~\bibnamefont {Pollack}},\ }\href
  {https://doi.org/10.1016/j.physletb.2020.135657} {\bibfield  {journal}
  {\bibinfo  {journal} {Phys. Lett. B}\ }\textbf {\bibinfo {volume} {808}},\
  \bibinfo {pages} {135657} (\bibinfo {year} {2020})},\ \Eprint
  {https://arxiv.org/abs/2004.12995} {arXiv:2004.12995 [gr-qc]} \BibitemShut
  {NoStop}%
\bibitem [{\citenamefont {Hennigar}\ \emph {et~al.}(2021)\citenamefont
  {Hennigar}, \citenamefont {Kubiznak},\ and\ \citenamefont
  {Mann}}]{Hennigar:2020drx}%
  \BibitemOpen
  \bibfield  {author} {\bibinfo {author} {\bibfnamefont {R.~A.}\ \bibnamefont
  {Hennigar}}, \bibinfo {author} {\bibfnamefont {D.}~\bibnamefont {Kubiznak}},\
  and\ \bibinfo {author} {\bibfnamefont {R.~B.}\ \bibnamefont {Mann}},\ }\href
  {https://doi.org/10.1088/1361-6382/abce48} {\bibfield  {journal} {\bibinfo
  {journal} {Class. Quant. Grav.}\ }\textbf {\bibinfo {volume} {38}},\ \bibinfo
  {pages} {03LT01} (\bibinfo {year} {2021})},\ \Eprint
  {https://arxiv.org/abs/2005.13732} {arXiv:2005.13732 [gr-qc]} \BibitemShut
  {NoStop}%
\bibitem [{\citenamefont {Aoki}\ \emph {et~al.}(2020)\citenamefont {Aoki},
  \citenamefont {Gorji},\ and\ \citenamefont {Mukohyama}}]{Aoki:2020lig}%
  \BibitemOpen
  \bibfield  {author} {\bibinfo {author} {\bibfnamefont {K.}~\bibnamefont
  {Aoki}}, \bibinfo {author} {\bibfnamefont {M.~A.}\ \bibnamefont {Gorji}},\
  and\ \bibinfo {author} {\bibfnamefont {S.}~\bibnamefont {Mukohyama}},\ }\href
  {https://doi.org/10.1016/j.physletb.2020.135843} {\bibfield  {journal}
  {\bibinfo  {journal} {Phys. Lett. B}\ }\textbf {\bibinfo {volume} {810}},\
  \bibinfo {pages} {135843} (\bibinfo {year} {2020})},\ \Eprint
  {https://arxiv.org/abs/2005.03859} {arXiv:2005.03859 [gr-qc]} \BibitemShut
  {NoStop}%
\bibitem [{\citenamefont {Cardoso}\ and\ \citenamefont
  {Lemos}(2001)}]{Cardoso:2001hn}%
  \BibitemOpen
  \bibfield  {author} {\bibinfo {author} {\bibfnamefont {V.}~\bibnamefont
  {Cardoso}}\ and\ \bibinfo {author} {\bibfnamefont {J.~P.~S.}\ \bibnamefont
  {Lemos}},\ }\href {https://doi.org/10.1103/PhysRevD.63.124015} {\bibfield
  {journal} {\bibinfo  {journal} {Phys. Rev. D}\ }\textbf {\bibinfo {volume}
  {63}},\ \bibinfo {pages} {124015} (\bibinfo {year} {2001})},\ \Eprint
  {https://arxiv.org/abs/gr-qc/0101052} {arXiv:gr-qc/0101052} \BibitemShut
  {NoStop}%
\bibitem [{\citenamefont {Konoplya}(2004)}]{Konoplya:2004ik}%
  \BibitemOpen
  \bibfield  {author} {\bibinfo {author} {\bibfnamefont {R.~A.}\ \bibnamefont
  {Konoplya}},\ }\href {https://doi.org/10.1103/PhysRevD.70.047503} {\bibfield
  {journal} {\bibinfo  {journal} {Phys. Rev. D}\ }\textbf {\bibinfo {volume}
  {70}},\ \bibinfo {pages} {047503} (\bibinfo {year} {2004})},\ \Eprint
  {https://arxiv.org/abs/hep-th/0406100} {arXiv:hep-th/0406100} \BibitemShut
  {NoStop}%
\bibitem [{\citenamefont {Fontana}(2023)}]{Fontana:2023dix}%
  \BibitemOpen
  \bibfield  {author} {\bibinfo {author} {\bibfnamefont {R.~D.~B.}\
  \bibnamefont {Fontana}},\ }\href@noop {} {\  (\bibinfo {year} {2023})},\
  \Eprint {https://arxiv.org/abs/2306.02504} {arXiv:2306.02504 [gr-qc]}
  \BibitemShut {NoStop}%
\bibitem [{\citenamefont {Birmingham}\ \emph {et~al.}(2002)\citenamefont
  {Birmingham}, \citenamefont {Sachs},\ and\ \citenamefont
  {Solodukhin}}]{Birmingham:2001pj}%
  \BibitemOpen
  \bibfield  {author} {\bibinfo {author} {\bibfnamefont {D.}~\bibnamefont
  {Birmingham}}, \bibinfo {author} {\bibfnamefont {I.}~\bibnamefont {Sachs}},\
  and\ \bibinfo {author} {\bibfnamefont {S.~N.}\ \bibnamefont {Solodukhin}},\
  }\href {https://doi.org/10.1103/PhysRevLett.88.151301} {\bibfield  {journal}
  {\bibinfo  {journal} {Phys. Rev. Lett.}\ }\textbf {\bibinfo {volume} {88}},\
  \bibinfo {pages} {151301} (\bibinfo {year} {2002})},\ \Eprint
  {https://arxiv.org/abs/hep-th/0112055} {arXiv:hep-th/0112055} \BibitemShut
  {NoStop}%
\bibitem [{\citenamefont {Banados}\ \emph {et~al.}(1992)\citenamefont
  {Banados}, \citenamefont {Teitelboim},\ and\ \citenamefont
  {Zanelli}}]{Banados:1992wn}%
  \BibitemOpen
  \bibfield  {author} {\bibinfo {author} {\bibfnamefont {M.}~\bibnamefont
  {Banados}}, \bibinfo {author} {\bibfnamefont {C.}~\bibnamefont
  {Teitelboim}},\ and\ \bibinfo {author} {\bibfnamefont {J.}~\bibnamefont
  {Zanelli}},\ }\href {https://doi.org/10.1103/PhysRevLett.69.1849} {\bibfield
  {journal} {\bibinfo  {journal} {Phys. Rev. Lett.}\ }\textbf {\bibinfo
  {volume} {69}},\ \bibinfo {pages} {1849} (\bibinfo {year} {1992})},\ \Eprint
  {https://arxiv.org/abs/hep-th/9204099} {arXiv:hep-th/9204099} \BibitemShut
  {NoStop}%
\bibitem [{\citenamefont {Gundlach}\ \emph {et~al.}(1994)\citenamefont
  {Gundlach}, \citenamefont {Price},\ and\ \citenamefont
  {Pullin}}]{Gundlach:1993tp}%
  \BibitemOpen
  \bibfield  {author} {\bibinfo {author} {\bibfnamefont {C.}~\bibnamefont
  {Gundlach}}, \bibinfo {author} {\bibfnamefont {R.~H.}\ \bibnamefont
  {Price}},\ and\ \bibinfo {author} {\bibfnamefont {J.}~\bibnamefont
  {Pullin}},\ }\href {https://doi.org/10.1103/PhysRevD.49.883} {\bibfield
  {journal} {\bibinfo  {journal} {Phys. Rev. D}\ }\textbf {\bibinfo {volume}
  {49}},\ \bibinfo {pages} {883} (\bibinfo {year} {1994})},\ \Eprint
  {https://arxiv.org/abs/gr-qc/9307009} {arXiv:gr-qc/9307009} \BibitemShut
  {NoStop}%
\bibitem [{\citenamefont {Konoplya}\ and\ \citenamefont
  {Fontana}(2008)}]{Konoplya:2007yy}%
  \BibitemOpen
  \bibfield  {author} {\bibinfo {author} {\bibfnamefont {R.~A.}\ \bibnamefont
  {Konoplya}}\ and\ \bibinfo {author} {\bibfnamefont {R.~D.~B.}\ \bibnamefont
  {Fontana}},\ }\href {https://doi.org/10.1016/j.physletb.2007.10.065}
  {\bibfield  {journal} {\bibinfo  {journal} {Phys. Lett. B}\ }\textbf
  {\bibinfo {volume} {659}},\ \bibinfo {pages} {375} (\bibinfo {year}
  {2008})},\ \Eprint {https://arxiv.org/abs/0707.1156} {arXiv:0707.1156
  [hep-th]} \BibitemShut {NoStop}%
\bibitem [{\citenamefont {Churilova}\ and\ \citenamefont
  {Stuchlik}(2020)}]{Churilova:2019cyt}%
  \BibitemOpen
  \bibfield  {author} {\bibinfo {author} {\bibfnamefont {M.~S.}\ \bibnamefont
  {Churilova}}\ and\ \bibinfo {author} {\bibfnamefont {Z.}~\bibnamefont
  {Stuchlik}},\ }\href {https://doi.org/10.1088/1361-6382/ab7717} {\bibfield
  {journal} {\bibinfo  {journal} {Class. Quant. Grav.}\ }\textbf {\bibinfo
  {volume} {37}},\ \bibinfo {pages} {075014} (\bibinfo {year} {2020})},\
  \Eprint {https://arxiv.org/abs/1911.11823} {arXiv:1911.11823 [gr-qc]}
  \BibitemShut {NoStop}%
\bibitem [{\citenamefont {Bronnikov}\ and\ \citenamefont
  {Konoplya}(2020)}]{Bronnikov:2019sbx}%
  \BibitemOpen
  \bibfield  {author} {\bibinfo {author} {\bibfnamefont {K.~A.}\ \bibnamefont
  {Bronnikov}}\ and\ \bibinfo {author} {\bibfnamefont {R.~A.}\ \bibnamefont
  {Konoplya}},\ }\href {https://doi.org/10.1103/PhysRevD.101.064004} {\bibfield
   {journal} {\bibinfo  {journal} {Phys. Rev. D}\ }\textbf {\bibinfo {volume}
  {101}},\ \bibinfo {pages} {064004} (\bibinfo {year} {2020})},\ \Eprint
  {https://arxiv.org/abs/1912.05315} {arXiv:1912.05315 [gr-qc]} \BibitemShut
  {NoStop}%
\bibitem [{\citenamefont {Konoplya}\ \emph {et~al.}(2019)\citenamefont
  {Konoplya}, \citenamefont {Zhidenko},\ and\ \citenamefont
  {Zinhailo}}]{Konoplya:2019hlu}%
  \BibitemOpen
  \bibfield  {author} {\bibinfo {author} {\bibfnamefont {R.~A.}\ \bibnamefont
  {Konoplya}}, \bibinfo {author} {\bibfnamefont {A.}~\bibnamefont {Zhidenko}},\
  and\ \bibinfo {author} {\bibfnamefont {A.~F.}\ \bibnamefont {Zinhailo}},\
  }\href {https://doi.org/10.1088/1361-6382/ab2e25} {\bibfield  {journal}
  {\bibinfo  {journal} {Class. Quant. Grav.}\ }\textbf {\bibinfo {volume}
  {36}},\ \bibinfo {pages} {155002} (\bibinfo {year} {2019})},\ \Eprint
  {https://arxiv.org/abs/1904.10333} {arXiv:1904.10333 [gr-qc]} \BibitemShut
  {NoStop}%
\bibitem [{\citenamefont {Matyjasek}\ and\ \citenamefont
  {Opala}(2017)}]{Matyjasek:2017psv}%
  \BibitemOpen
  \bibfield  {author} {\bibinfo {author} {\bibfnamefont {J.}~\bibnamefont
  {Matyjasek}}\ and\ \bibinfo {author} {\bibfnamefont {M.}~\bibnamefont
  {Opala}},\ }\href {https://doi.org/10.1103/PhysRevD.96.024011} {\bibfield
  {journal} {\bibinfo  {journal} {Phys. Rev. D}\ }\textbf {\bibinfo {volume}
  {96}},\ \bibinfo {pages} {024011} (\bibinfo {year} {2017})},\ \Eprint
  {https://arxiv.org/abs/1704.00361} {arXiv:1704.00361 [gr-qc]} \BibitemShut
  {NoStop}%
\bibitem [{\citenamefont {Konoplya}(2003)}]{Konoplya:2003ii}%
  \BibitemOpen
  \bibfield  {author} {\bibinfo {author} {\bibfnamefont {R.~A.}\ \bibnamefont
  {Konoplya}},\ }\href {https://doi.org/10.1103/PhysRevD.68.024018} {\bibfield
  {journal} {\bibinfo  {journal} {Phys. Rev. D}\ }\textbf {\bibinfo {volume}
  {68}},\ \bibinfo {pages} {024018} (\bibinfo {year} {2003})},\ \Eprint
  {https://arxiv.org/abs/gr-qc/0303052} {arXiv:gr-qc/0303052} \BibitemShut
  {NoStop}%
\bibitem [{\citenamefont {Konoplya}\ \emph {et~al.}(2008)\citenamefont
  {Konoplya}, \citenamefont {Murata}, \citenamefont {Soda},\ and\ \citenamefont
  {Zhidenko}}]{Konoplya:2008yy}%
  \BibitemOpen
  \bibfield  {author} {\bibinfo {author} {\bibfnamefont {R.~A.}\ \bibnamefont
  {Konoplya}}, \bibinfo {author} {\bibfnamefont {K.}~\bibnamefont {Murata}},
  \bibinfo {author} {\bibfnamefont {J.}~\bibnamefont {Soda}},\ and\ \bibinfo
  {author} {\bibfnamefont {A.}~\bibnamefont {Zhidenko}},\ }\href
  {https://doi.org/10.1103/PhysRevD.78.084012} {\bibfield  {journal} {\bibinfo
  {journal} {Phys. Rev. D}\ }\textbf {\bibinfo {volume} {78}},\ \bibinfo
  {pages} {084012} (\bibinfo {year} {2008})},\ \Eprint
  {https://arxiv.org/abs/0807.1897} {arXiv:0807.1897 [hep-th]} \BibitemShut
  {NoStop}%
\bibitem [{\citenamefont {Zhu}\ \emph {et~al.}(2014)\citenamefont {Zhu},
  \citenamefont {Zhang}, \citenamefont {Pellicer}, \citenamefont {Wang},\ and\
  \citenamefont {Abdalla}}]{Zhu:2014sya}%
  \BibitemOpen
  \bibfield  {author} {\bibinfo {author} {\bibfnamefont {Z.}~\bibnamefont
  {Zhu}}, \bibinfo {author} {\bibfnamefont {S.-J.}\ \bibnamefont {Zhang}},
  \bibinfo {author} {\bibfnamefont {C.~E.}\ \bibnamefont {Pellicer}}, \bibinfo
  {author} {\bibfnamefont {B.}~\bibnamefont {Wang}},\ and\ \bibinfo {author}
  {\bibfnamefont {E.}~\bibnamefont {Abdalla}},\ }\href
  {https://doi.org/10.1103/PhysRevD.90.044042} {\bibfield  {journal} {\bibinfo
  {journal} {Phys. Rev. D}\ }\textbf {\bibinfo {volume} {90}},\ \bibinfo
  {pages} {044042} (\bibinfo {year} {2014})},\ \bibinfo {note} {[Addendum:
  Phys.Rev.D 90, 049904 (2014)]},\ \Eprint {https://arxiv.org/abs/1405.4931}
  {arXiv:1405.4931 [hep-th]} \BibitemShut {NoStop}%
\bibitem [{\citenamefont {Konoplya}\ and\ \citenamefont
  {Zhidenko}(2014)}]{Konoplya:2014lha}%
  \BibitemOpen
  \bibfield  {author} {\bibinfo {author} {\bibfnamefont {R.~A.}\ \bibnamefont
  {Konoplya}}\ and\ \bibinfo {author} {\bibfnamefont {A.}~\bibnamefont
  {Zhidenko}},\ }\href {https://doi.org/10.1103/PhysRevD.90.064048} {\bibfield
  {journal} {\bibinfo  {journal} {Phys. Rev. D}\ }\textbf {\bibinfo {volume}
  {90}},\ \bibinfo {pages} {064048} (\bibinfo {year} {2014})},\ \Eprint
  {https://arxiv.org/abs/1406.0019} {arXiv:1406.0019 [hep-th]} \BibitemShut
  {NoStop}%
\end{thebibliography}%

\end{document}